\documentclass[twocolumn,showpacs,amsmath,amssymb]{revtex4}

\usepackage{amsmath} %
\usepackage{graphicx}
\usepackage{amsfonts}
\usepackage{dcolumn}
\usepackage{bm}

\newcommand\e{\epsilon}

\begin{document}
\title{Orientation Sensitive Nonlinear Growth of Graphene: A Geometry-determined Epitaxial Growth Mechanism}

\author{Huijun Jiang}
\author{Ping Wu}
\author{Zhonghuai Hou}
\thanks{Corresponding author. E-mail: hzhlj@ustc.edu.cn}
\author{Zhenyu Li}
\thanks{Corresponding author. E-mail: zyli@ustc.edu.cn}
\author{Jinlong Yang}
\affiliation{Department of Chemical Physics \& Hefei National Laboratory for Physical Sciences at Microscales, University of Science and Technology of China, Hefei, Anhui 230026, China}
\date{\today}

\begin{abstract}
Although the corresponding carbon-metal interactions can be very different, a similar nonlinear growth behavior of graphene has been observed for different metal substrates. To understand this interesting experimental observation, a multiscale $\lq\lq$standing-on-the-front" kinetic Monte Carlo study is performed. An extraordinary robust geometry effect is identified, which solely determines the growth kinetics and makes the details of carbon-metal interaction not relevant at all. Based on such a geometry-determined mechanism, epitaxial growth behavior of graphene can be easily predicted in many cases. As an example, an orientation-sensitive growth kinetics of graphene on Ir(111) surface has been studied. Our results demonstrate that lattice mismatch pattern at the atomic level  plays an important role for macroscopic epitaxial growth.
\end{abstract}

\maketitle
Due to  its excellent electronic, mechanical, thermal, and optical performance, graphene has drawn great attentions of physicists, chemists, and material scientists\cite{Sci04000666}. Among several ways to produce graphene\cite{Sci04000666,CSR10000228}, epitaxial growth on metal surfaces is of particular importance because it can generate large size graphene sample of high quality \cite{NtM08000406,NaL08000565,Nat09000706,ACS11003385,Sci09001312}.  Plenty of nontrivial growth behaviors have been revealed in experiments\cite{NtM08000406,PRL08107602,NaL08000030,PRB07075429,PRL09056808,PRL09166101,JPCC1210557,NJP08093026}. Of particular interest, the growth rate is found to be a quintic function of carbon monomer concentration on the main growth orientation (R0 orientation) of Ir(111) surface, which suggests a growth mechanism with the attachment of five-atom clusters\cite{NJP08093026}. The same nonlinearity is also reported on Ru(0001) surface\cite{NJP09063046}, indicating a common growth mechanism shared by different metal surfaces. This is counterintuitive, since the carbon-metal interaction which determines the graphene growth is different for different substrates. What's more, graphene growth on the same Ir(111) surface but for another orientation R30 has different growth rate and nonlinear dependence \cite{PRB09085430,NJP09063046}. The similar growth kinetics on different substrates and different growth kinetics on the same substrate found in experiment present a very attractive mystery.

While there are a variety of experimental works in this field, theoretical studies to reveal the growth kinetics are rare. Recently, by assuming that graphene islands grow homogeneously via the attachment of five-atom carbon ($C_5$) clusters \cite{NaL11002092}, a rate theory was developed  to produce a quantitative account of the measured time-dependent carbon adatom density. Nevertheless, this model was mainly phenomenological and provide little atomic mechanisms. On the other hand,  the large time scale discrepancies between the attaching/detaching events at the atomic level and the growth behaviors at the macroscopic level, as well as those between the growth events involving different carbon clusters, make it hardly possible for brute-force simulations. To bridges the gap between lattice mismatch pattern at the atomic level and macroscopic experimental growth kinetics, we have proposed a $\lq\lq$standing-on-the-front" kinetic Monte Carlo (SOF-kMC) approach combining with density functional theory (DFT) calculations to perform multiscale simulations of the epitaxial growth of graphene in our earlier work \cite{JAC12006045}. There, we have performed extensive DFT calculations to get the atomic details of different carbon clusters on the surface, and mainly focused on the growth behavior along R0 orientation on Ir surface. The obtained growth rate shows nonlinear growth behavior in very good agreements with the experiments. By physical intuition, we proposed therein that lattice mismatch should play an important role for the nonlinear growth behavior. Although the mentioned work has made an important step toward the understanding of graphene growth kinetics, more interesting questions also arise. For instance, why the growth exponent is about 5 (experimental value is slightly larger than 5), but not 4 or 6? Is the picture that only C5 successfully lead to graphene growth right or not? Why similar nonlinear growth behavior can be observed on different metal surfaces? Why the nonlinear behavior is so sensitive to the growth orientation? To answer these questions, a fundamental understanding of the epitaxial growth kinetics, relating the macroscopic growth behaviors to the atomic details, is very demanded.

\begin{figure}
\begin{center}
\includegraphics[width=1.0\columnwidth] {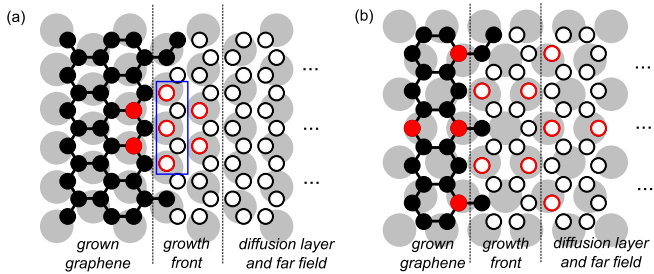}
\end{center}
\caption{Schematic diagrams of lattice mismatch pattern on (a) R0 and (b) R30 orientations of Ir(001). Gray, black and red circles represent the substrate atoms, ES and DS, respectively.
Solid(open) symbols stands for occupied (unoccupied) sites.} \label{fig:ltc}
\end{figure}

In the present paper, we apply the SOF-kMC approach combined with DFT calculations to address this problem. Generally, {\it lattice mismatch} will result in specific heterogeneity for graphene growth \cite{PRB07075429,JAC12006045}. Taking Ir(111) surface as a paradigm as shown in Fig.\ref{fig:ltc}, those \emph{difficult sites} (DS) nearly on top of the substrate atoms are hard to be attached by carbon atoms due to their high energy barriers of attachment, while other \emph{easy sites} (ES) are favorable for cabon atom attachment. With such a simplification from Morie pattern to a DS-ES model, we show that our multiscale approach can well reproduce the quintic growth kinetics observed experimentally for the R0 orientation. The underlying atomic mechanisms of this nontrivial growth kinetics, which can be revealed by detailed analysis of the involved growth events, are shown to be rather robust to the energy parameters associated with DS. Such a robustness may explain why similar nonlinear growth kinetics can be founded on different metal substrates. Furthermore, studies on R30 orientation reveal that the growth mechanism is much sensitive to the distribution of DS which is solely determined by the geometry of mismatched lattice. These findings bring us to a geometry-determined epitaxial growth mechanism of graphene on metal surfaces with lattice mismatch.

The starting point of our SOF-kMC model is realizing that the growth kinetics are determined by  the attachment and detachment processes of different  carbon clusters at a well-defined  growth front. The growth rate can be readily calculated if the incoming fluxes of all carbon clusters to the front and the involved rates of the attachment/detachment processes are available. In this regard, the whole surface lattice can be divided into four regions: The grown graphene sheet, growth front, diffusion layer, and far field, as illustrated briefly in Fig.\ref{fig:ltc} and  in more details in Fig.(S1) of the supplemental information (SI). In the far field, $C_i$  clusters are assumed to be in equilibrium with each other, which allows us to estimate their populations on the surface according to $iC_1\rightleftharpoons C_i$ with energies of each carbon species available. These clusters diffuse across the diffusion layer to the growth front, where they attach to the grown graphene with rate $k_i^a$, or detach from it with rate $k_i^d$. All of the energy parameters including $\e_{i,D(E)}^{a(d)}$ associated with the attaching(detaching) events of $C_i$ clusters on DS(ES) are all obtained from DFT calculations. Taking into account the clusters up to $i=6$ and the lattice mismatch induced heterogeneity, one may then build up a kinetic model with 24 key events taking place on the growth front, namely, the attachment and detachment of $C_{i=1,\dots,6}$ associated with DS and ES. Conventionally, a standard $\lq\lq$event-list" kMC algorithm  is ready for the simulation, which runs the dynamics by randomly determining what the next event is and when it will happen\cite{JPC77002340}. However, DFT calculations showed that there are very large(up to more than 10 orders of magnitude) discrepancies among the rates of these events, which render the direct kMC simulations very expensive. To overcome this difficulty, we have used  a nested kMC algorithm, which finally makes the simulation of graphene growth a tractable problem and facilitates the calculation of the growth rate $R_G$.  More details of the SOF-kMC approach are described in the SI.

To validate the SOF-kMC approach, we have simulated the dependence of $R_G$ on $n_1$, the population of $C_1$, for graphene growth on the R0 orientation on the Ir surface \cite{JAC12006045}. The obtained curve can be very well fitted by a nonlinear growing function, $R_G \sim a n_1^{\gamma} + b$, with the exponent $\gamma \simeq 5.25\pm0.02$ and $a, b$ two constants. One notes that the value $5.25$ agrees very well with the experiment\cite{NJP09063046}, which confirms the validity of our method. It is worthy to emphasize here that the growth exponent $\gamma$ is not exactly 5, but slightly larger. This suggests that the picture proposed in the literatures\cite{NJP08093026,NaL11002092} that only $C_5$ clusters can effectively contribute to the front growth was not exactly right. Using our approach, it is feasible to perform a detail analysis about the key events that contribute to the front growth. To this end, we have counted the numbers of each event that really happened in simulation of the front movement, as shown in Table \ref{tab:GPR0}. Clearly, growth over ES is dominated by $C_1$ attachment. On the contrary,  only large clusters ($C_4$, $C_5$ and $C_6$) can significantly attach to DS. The nonlinear growth behavior suggests that, it is such DS-attaching events involving these relatively large clusters that control the growth process.

However, the results shown in Table \ref{tab:GPR0} indicate that $C_4$ clusters attach more frequently to DS than $C_5$ and $C_6$, which would suggest a nonlinear growth exponent $4<\gamma<5$. The simulation value $\gamma=5.25$ seems to imply that $C_4$ actually  contributes little to the front growth. To elucidate this point, we have traced the carbon atoms attached onto DS. We find that although $C_4$ clusters can attach to DS easily, almost all of them will detach via $C_2$ clusters such that the net contribution to the front growth is negligibly small.  The attached $C_5$ clusters may also detach via small clusters, but an apparently nonvanishing amount of $C_5$ clusters will stay and thus provide a net contribution to the front growth. There is also a net contribution from $C_6$ clusters, but it is much less significant compared to $C_5$. In Table \ref{tab:RPR0}, the detaching probabilities of attached $C_4$, $C_5$ and $C_6$ via small clusters, as well as their net attachment probabilities are listed. Clearly, only $C_5$ and $C_6$ can stay on DS stably  and contribute to the front growth, and $C_5$ is significantly more important than $C_6$. Thus, the highly nonlinear growth behavior is a cooperative effect associated with the attachments of several  large carbon clusters, rather than with a single species $C_5$.

\begin{table}
\caption{Relative probabilities of growth events on R0 orientation for $n_1=0.01ML$. $\lq$A' and $\lq$D' in brackets stand for attachment and detachment events, respectively.}
\label{tab:GPR0}
\renewcommand{\baselinestretch}{1.0}\normalsize
\setlength{\tabcolsep}{0.01\columnwidth}
\begin{center}
\begin{tabular}{c|c|c|c|c}
\hline \hline
& ES(A)&ES(D) & DS(A)& DS(D)\\
\hline
$C_1$& $0.342$ & $0.219$ & $\sim$ & $1.79\times10^{-3}$ \\
\hline
$C_2$& $1.10\times10^{-5}$ & $3.27\times10^{-3}$ & $\sim$ & $0.288$ \\
\hline
$C_3$& $\sim$ & $\sim$ &$\sim$ & $5.38\times10^{-4}$\\
\hline
$C_4$& $\sim$ & $\sim$ &$0.139$ & $\sim$ \\
\hline
$C_5$& $\sim$ & $\sim$ &$7.51 \times10^{-3}$ & $\sim$ \\
\hline
$C_6$& $\sim$ & $\sim$ &$1.65\times10^{-5}$ & $\sim$ \\
\hline \hline
\end{tabular}
\end{center}
\begin{flushleft}
$\sim$: negligibly small.
\end{flushleft}
\end{table}

One should note that the growth process is rate-limited by DS, i.e., only when all the DS are filled stably, can the front move forward.  The above analysis reveals that only $C_5$ and $C_6$ clusters can effectively lead to front growth, i.e., stably fill the DS, despite their low populations on the surface.  To unravel this mystery, it is instructive  to turn to the mismatch pattern shown in Fig.\ref{fig:ltc}(a).  For the zigzag-shape front enclosed by the rectangle, three nearest DS
occupy two adjacent hexagons. Small clusters can hardly fill these three DS simultaneously, thus they cannot be stably attached and will detach quickly before the next coming species.  A $C_4$ cluster will fill two adjacent DS and complete one hexagon, which makes the attaching events of $C_4$ on DS possible as shown in Table \ref{tab:GPR0}. However, the third unfilled DS may destabilize the front and  these attached $C_4$ clusters will detach via smaller clusters with very small energy barrieres as illustrated in Table \ref{tab:RPR0}. On the contrary, a $C_5$ or $C_6$ cluster can cover these three DS via a single attaching event, such that the resulted conformation is rather stable because the two adjacent hexagons are both completed and hence the front will grow successfully. The contribution of $C_6$ is much smaller simply because its population is much smaller on the surface. Therefore, it is the  distribution of DS over the surface that determines the dominant DS-attaching species, indicating a solely geometry effect (As demonstrated in Ref.\cite{PRB09085430}, the locations of DS can be derived by using simple geometric rules involving periodic and quasiperiodic structural motifs). In a word, the observed nonlinear growth behavior of graphene actually bears a geometry-determined atomic mechanism.

\begin{table}
\caption{Relative probabilities of DS events  during the growth process on R0 orientation for $n_1=0.01ML$.}
\label{tab:RPR0}
\renewcommand{\baselinestretch}{1.0}\normalsize
\setlength{\tabcolsep}{0.01\columnwidth}
\begin{center}
\begin{tabular}{c|c|c|c|c}
\hline \hline
\multicolumn{2}{c|} {Species}& $C_4$ &$C_5$ &$C_6$\\
\hline
\multicolumn{2}{c|} {Attachment}& $0.316$ & $0.0171$ &$6.58\times10^{-5}$\\
\hline
& $C_1$ & $\sim$ & $4.52\times10^{-3}$ & $\sim$\\
Detachment via& $C_2$ & $0.632$ & $0.0287$ & $1.39\times10^{-4}$\\
&$C_3$& $\sim$ & $1.11\times10^{-3}$ & $\sim$\\
\hline
\multicolumn{2}{c|} {Net contribution}& $\sim$ & $4.01\times10^{-3}$ & $1.94\times10^{-5}$\\
\hline \hline
\end{tabular}
\end{center}
\begin{flushleft}
$\sim$: negligibly small.
\end{flushleft}
\end{table}

\begin{figure}
\begin{center}
\includegraphics[width=0.8\columnwidth] {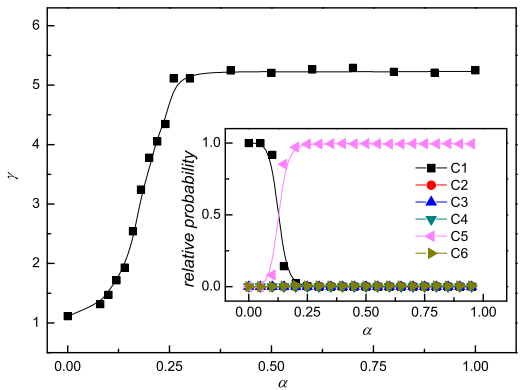}
\end{center}
\caption{The dependence of growth exponent $\gamma$ on heterogeneity level $\alpha$ for R0 orientation. Inset: Net contribution of different carbon species (in relative probability).} \label{fig:NLR0}
\end{figure}

If such a  geometry-determined mechanism is right,  one can expect that the nonlinear growth bevavior  should be robust to the heterogeneity level, which can be measured by the discrepancies between the energy parameters associated with DS and those with ES:  $\Delta \e_i^{a(d)}=\e_{i,D}^{a(d)}-\e_{i,E}^{a(d)}$.  Although these values are fixed for the growth on Ir surface, which can be noted as $\Delta \e_{i0}^{a(d)}$, it is convenient for us to treat them as variable parameters in our simulation framework. This should be of particular importance, on the one hand, to get more new insights into the underlying mechanism of the lattice mismatch induced nonlinear growth, and on the other hand, to give predictions about the growth behaviors on other metal surfaces with similar lattice mismatch pattern but with different heterogeneity level. The dependence of the growth exponent $\gamma$ on $\alpha=\Delta \e_i^{a(d)}/\Delta \e_{i0}^{a(d)}$ is shown in Fig.\ref{fig:NLR0}. For $\alpha=0$, the surface is essentially homogeneous and surely the growth process would be determined by  $C_1$ which corresponds to $\gamma=1$. With increasing $\alpha$, lattice mismatch takes effect and carbon clusters are required to fill DS, which leads to nonlinear growth with $\gamma>1$. Remarkably, two distinct regimes are observed with increasing $\alpha$: For $\alpha$ less than a certain threshold value $\alpha_c$, the exponent $\gamma$ is rather sensitive to the change of $\alpha$; while for  $\alpha>\alpha_c$,  $\gamma$ saturates to a nearly constant value same as that for $\alpha=1$. This validates that, if  lattice mismatch induced heterogeneity is strong enough, the growth behavior is indeed robust to it and solely determined by the geometry.

We have also analyzed the net contributions (probabilities) of different $C_i$ species to the front growth as functions of $\alpha$. As shown in the inset of Fig.\ref{fig:NLR0}, a narrow transition window exists around $\alpha_c$. For small $\alpha$, the front growth is dominated by $C_1$, the net probability of which decreases sharply within the window. On the contrary, the contribution of $C_5$ increases sharply in the window and becomes dominate for $\alpha>\alpha_c$. Similar to the case of $\alpha=1$, as demonstrated in Table \ref{tab:GPR0}, the contributions of $C_i$ with $(i=2,3,4)$ to the front growth is negligible. The fact that only $C_5$ and $C_6$ (much smaller) can contribute to the front growth for relatively large $\alpha$ further demonstrate the validity of geometry-determinied mechanism.  We note here that the robustness of $\gamma$ to the change of $\e_{i,D}^{a(d)}$ for each single growth event has also been tested (see the SI).

\begin{figure}
\begin{center}
\includegraphics[width=0.7\columnwidth] {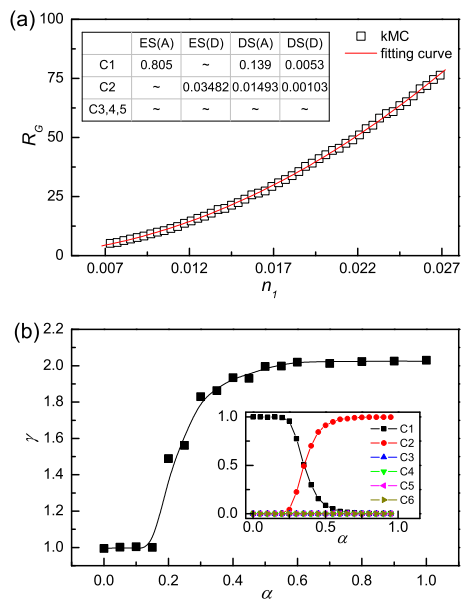}
\end{center}
\caption{(a) Dependence of  $R_G$ on $n_1$ (in $ML$) on R30 orientation. Inset: Relative probabilities of growth events. (b) Growth exponent $\gamma$ as a function of heterogeneity level $\alpha$ for R30 orientation. Inset: Net contribution of different carbon species (in relative probability).} \label{fig:rateR30}
\end{figure}

The above geometry-dependent picture raises an interesting question: How the growth kinetics would change if the graphene grows on another orientation, e.g., R30?  As shown in Fig.1(b), the mismatch pattern along R30 is quite different from that along R0, say, the DS are separated more apart from each other. To address this issue, we have performed similar simulations by using specific energy parameters $\e_{i,D(E)}^{a(d)}$ for the R30 direction obtained by DFT calculations. The dependence of $R_G$ on $n_1$ is shown in  Fig.\ref{fig:rateR30}(a), which can be well fitted by $R_G = a n_1^\gamma + b$ with an exponent $\gamma=2.01 \pm 0.02 $. The inset presents the relative probability of different growth events. Clearly, DS can now be filled successfully by $C_2$ clusters, which dominates the growth on R30. Contrary to the R0 case, contributions of the carbon clusters with $i\ge 3$ are not observable. We have also investigated how the growth kinetics depends on the heterogeneity level. The results similar to Fig.3 are shown in Fig.\ref{fig:rateR30}(b), where the exponent $\gamma$ is drawn as a function of $\alpha$. There is also an increasing of $\gamma$ with $\alpha$, but now $\gamma$ saturates to a much smaller value 2.0 compared to $\gamma=5.25$ in the R0 case. As shown in the inset, only $C_1$ and $C_2$ matter for this orientation, while the former dominates for small $\alpha$ and the latter dominates for large $\alpha$.

The observed exponent $\gamma \simeq 2$ for R30 as well as its robustness against the parameter $\alpha$ can also be elucidated by the same geometry-determined mechanism as that for R0.  As shown in Fig.1(b), DS on the front are separated from each other with at least three ES in between. Clearly, DS in the growth front can be filled by $C_2$ clusters successfully. When one DS is filled, the hexagon it belongs to is completed and stable which is not influenced by the other DS nearby. The front thus moves forward when the DS are all filled by $C_2$ species. The key difference between R0 and R30 directions is the correlation between adjacent DS, which is strong in the former and weak in the latter. Note that this correlation is simply determined by the mismatch geometry.

In summary, a geometry determined epitaxial growth mechanism of graphene on metal surface with lattice mismatch has been revealed in a general statistical mechanics framework by using a multiscale SOF-kMC approach. When sites difficult for monomer adsorption exist, growth kinetics can be well predicted simply by checking the mismatch pattern. We believe that our finding can inspire more experimental work and open new perspectives for theoretical studies on epitaxial growth kinetics.

We are grateful to Prof. Zhenyu Zhang for helpful discussions. This work is supported by MOST (2011CB921404), by NSFC (21125313,  20933006, 91027012, 21173202, 21121003, and 21222304), by CUSF (WK2340000011), by CAS(KJCX2-YWW22, XDB01020300), and by USTC-SCC, SCCAS, and Shanghai Supercomputer Centers.


\end{document}